\documentclass[final,3p,times,twocolumn]{elsarticle}

\usepackage{amssymb}
\usepackage{hyperref}

\usepackage{color}

\journal{Journal of Molecular Spectroscopy}

\begin{document}

\begin{frontmatter}



\title{Photoassociation of ultracold molecules near a Feshbach resonance as a probe of the electron-proton mass ratio variation}


\author{Marko Gacesa and Robin C\^ot\'e}


\address{University of Connecticut, Department of Physics, Storrs, CT 06269-3046, USA}

\begin{abstract}
We show that the photoassociation (PA) rate of ultracold diatomic molecules in the vicinity of a Feshbach resonance could be used to probe variations of the electron-to-proton mass ratio $\beta = m_e/m_p$, a quantity related to other fundamental constants via the $\Lambda_{\mathrm{QCD}}$ scale. The PA rate exhibits two features near a Feshbach resonance, a minimum and a maximum, which are very sensitive to the details of the interactions and the exact mass of the system. The effect and detection threshold are illustrated in the formation rates of ultracold Li$_2$ and LiNa molecules by performing coupled-channel calculations in an external magnetic field. We find that the PA rate is particularly sensitive near narrow Feshbach resonances in heavy molecules, where it might be possible to detect relative variability of $\beta$ on the order of $10^{-15}-10^{-16}$. We also use a two-channel model to derive a proportionality relation between the variation of the PA rate and $\beta$ applicable to diatomic molecules.
\end{abstract}


\end{frontmatter}







\section{Introduction}

Quasar absorption spectra hint that the fundamental constants may have changed over the history of the Universe \cite{2003MNRAS.345..609M,2008EPJST.163..159F,2011PhRvL.107s1101W,2012MNRAS.422.3370K,2012arXiv1202.6365K,2012A&A...542A.118B,2013arXiv1305.1884M}. If proved correct, there would be a profound impact on the fundamental laws of physics. Perhaps most importantly, the equivalence principle in general relativity would be violated, leading to a change of our understanding of cosmology \cite{will1993theory}.
Modern theories aiming to unify the four fundamental forces often allow temporal and spatial variations of the fine structure constant $\alpha=e^2/\hbar c$ and the electron-proton mass ratio $\beta = m_e/m_p$ due to the dramatic changes in the Universe's structure during its evolution. Grand unification theories also suggest more than one mechanism to make all coupling constants and elementary particle masses time- and space-dependent. A detailed review of the connection between the fundamental constants and their significance in different theoretical frameworks can be found in Ref. \cite{2011LRR....14....2U}.

In addition to astrophysical observations, limits on detection of the space-time variability of fundamental constants have been established from the Oklo natural nuclear reactor \cite{2006PhRvL..96o1101R}, Big Bang Nucleosynthesis \cite{2004PhRvD..69f3506D}, cosmic microwave background \cite{2010A&A...517A..62L}, meteorites \cite{PhysRevLett.91.261101}, and precision laboratory measurements using optical and microwave atomic fountain clocks \cite{PhysRevLett.74.3511,PhysRevLett.90.150801,PhysRevLett.98.070801,2008Sci...319.1808R}. A good summary of the present day values is given in Refs.  \cite{2008EPJST.163..159F,2011PhRvL.107s1101W,2011PThPh.126..993C}. 
So far these experiments have not detected a variation of any of the fundamental constants. However, this neither confirms nor contradicts the quasar spectra observations since they probe very different epochs in the evolution of the Universe.

Recent developments in ultracold atomic and molecular physics offer new possibilities for precision measurement experiments \cite{2009NJPh...11e5048C}. On one hand, atomic and molecular spectra depend on $\alpha$ through relativistic corrections which are observable as the fine structure, Lamb shift, hyperfine structure, and other small perturbations. On the other hand, molecular bound levels and certain types of molecular couplings depend on the mass, making phenomena depending on them (such as Feshbach resonances in ultracold atomic samples \cite{PhysRevLett.96.230801}) particularly well adapted to investigate the relative variation of dimensionless electron-proton mass ratio ${\delta\beta}/{\beta}$ over time. This can be understood as molecules are bound by the electronic interaction (making molecular interaction potential electronic in origin), while the mass of the individual atoms is mostly baryonic. Both scattering and spectroscopic properties of ultracold molecules, such as the scattering length or position of the last vibrational energy level, are very sensitive to $\beta$. Particularly suitable atomic and molecular systems have quasi-degenerate levels caused by the coupling terms of different nature. 
A number of methods to measure the variation of $\beta$ in such systems have been proposed  \cite{2009NJPh...11e5048C,PhysRevA.59.230,2008PhRvL.100d3202D,2008PhRvL.100d3201Z,PhysRevA.83.052706,PhysRevLett.106.210802,2012Natur.482...45Y,2012PhRvL.109g0802B,2013PhRvA..87e2509J}. The suggested experiments may be able to detect a time variation on the order of  $\dot{\beta}/ \beta \sim 10^{-15}$/year or better \cite{2011LRR....14....2U,2006PhRvL..96o1101R,2008PhRvL.100d3202D}, independently of the results obtained by the atomic clocks.

Chin and Flambaum have studied the variation of $\beta$ near a Feshbach resonance and derived a proportionality relation between the relative variations of the zero-energy atom-atom scattering length \cite{PhysRevLett.96.230801} and electron-proton mass ratio. The proportionality factor between the two quantities can be very large for suitable systems, implying that an indirect method for detecting small changes in $\beta$ could be based on measuring the relative variation of the scattering length near a Feshbach resonance. This approach can be applied to any diatomic system as long as Feshbach resonances can be found or induced.

Motivated by the result of Ref. \cite{PhysRevLett.96.230801} and growing interest in a search for variability of fundamental constants, we propose a new approach to accurately measure temporal variation of $\beta$ using photoassociative Feshbach spectroscopy in an ultracold gas. Photoassociative spectroscopy is an accurate and robust technique that has been used successfully to obtain high resolution spectra of molecular states in optically cooled gases \cite{PhysRevLett.58.2420,1999JMoSp.195..194S}, such as two-photon spectrocopy \cite{Cote-Killian-Sr_2}. It also allowed studies of the scattering length between atoms \cite{Cote-PRL-95,Cote-PRA-96,Cote-PRA-97}, and of the formation of ultracold molecules in their ground electronic states \cite{Cote-CPL-mol,Cote-JMS-mol,Cote-PRA-LiH}. 
It has been predicted theoretically \cite{2008PhRvL.101e3201P,PhysRevA.88.063418} and observed in experiments \cite{PhysRevLett.101.060406,2011PhRvA..84d3614M} that the PA rate of diatomic molecules near a magnetic Feshbach resonance can be enhanced by several orders of magnitude. The enhancement is caused by a large increase in the short-range amplitude of the open-channel component of the scattering wave function, leading to higher probability density and increased dipole transition rates between the scattering state and a target excited state at small internuclear separation.

In this article, we demonstrate the enhanced sensitivity of two features in the PA rate (a minimum and a maximum) near a Feshbach resonance to the variability of $\beta$, should such variation exist in Nature.
Our calculations suggest that an even larger proportionality factor than predicted in Ref. \cite{PhysRevLett.96.230801} can be expected between these quantities in suitable diatomic systems. We investigate in more detail the enhancement factor at the PA minimum, which is particularly interesting for potential experimental realizations as its position does not depend on the laser intensity \cite{2009NJPh...11e5047P,0953-4075-42-19-195202}, nor suffers from the perturbations present in the strongly interacting regime associated with the very large scattering length at the resonance.

\section{Theory}

The PA rate coefficient $K_b$, for a molecule photoassociated in a bound ro-vibrational level $b=(v,j)$ of a target electronic state from a pair of colliding atoms, can be expressed as $K_b = \langle v_{\mathrm{rel}} \sigma_b \rangle$, where $v_{\mathrm{rel}}$ is the relative velocity of the approaching atoms and $\sigma_b$ is the PA cross section for the corresponding bound level $b$.
Here, the brackets indicate averaging over the relative velocity distribution of the colliding atoms. If a Maxwell-Boltzman distribution of velocities  characterized by the temperature $T$ is assumed, the PA rate coefficient becomes \cite{1994PhRvL..73.1352N,1998PhRvA..58..498C}. 
\begin{equation}
  K_b(T) = \frac{k_B T}{h Q_T} \int_{0}^\infty{|S_b(\varepsilon,\ell,\omega)|^2 
                e^{-\varepsilon/{k_B T}}\frac{d\varepsilon}{k_B T}} , 
  \label{rate1}
\end{equation}
where $Q_T=(2\pi\mu k_B T/h^2)^{3/2}$ is the translational partition function and $\varepsilon=\hbar^2 k^2 / 2 \mu$ is the asymptotic kinetic energy for the colliding atom pair of reduced mass $\mu$, $\hbar \omega$ is the photon energy, $k_B$ is Boltzmann constant, $S_b(\varepsilon,\ell,\omega)$ is the $S$-matrix element for the PA process, and $\ell$ is the quantum number describing the relative angular momentum of the colliding atoms.
Near an isolated bound level $b$, the $S$-matrix has the form  \cite{1994PhRvL..73.1352N} 
\begin{equation}
  |S_b(\varepsilon,\ell,\omega)|^2 = \frac{\gamma_b \gamma_s}{(\varepsilon-\Delta_b)^2 + 
                           \frac{1}{4}(\gamma_s + \gamma_b)^2} ,
  \label{Sv}
\end{equation}
where $\Delta_b = E_b - \hbar \omega$ is the photon frequency detuning relative to the ro-vibrational level $b=(v,j)$ of the target molecular state of energy $E_b$, $\gamma_b$ is its natural linewidth, and $\gamma_s(\varepsilon,\ell)$ is the stimulated linewidth resulting from the laser coupling. Here, all other possible decay processes, such as predissociation, ionization, {\it etc}., are neglected. We note that in the unitarity limit, where $|S|^2=1$, the maximum value of the PA rate coefficient (\ref{rate1}) is given by
\begin{equation}
K^{\rm max}_b = \frac{k_B T}{h Q_T} = \frac{h^2}{(2\pi\mu)^{3/2}} \frac{1}{\sqrt{k_B T}} .
\label{eq:unitarity-limit}
\end{equation}

For low PA laser intensities, where multi-photon or ionization processes can be neglected, $\gamma_s$ can be obtained using the Fermi's golden rule, resulting in \cite{1994PhRvL..73.1352N,1998PhRvA..58..498C}
\begin{equation}
 \gamma_s(\varepsilon,\ell) = \frac{\pi I}{\epsilon_0 c} |\langle b | D(R)|\varepsilon, \ell \rangle|^2 ,
 \label{eq:gammas1}
\end{equation}
where $\epsilon_0$ and $c$ are the vacuum permittivity and speed of light, respectively, $I$ is the PA laser intensity, $|\varepsilon, \ell \rangle$ and $|b \rangle$ are the energy-normalized initial continuum state and the final bound ro-vibrational state, respectively, $D(R)$ is the transition dipole moment between these states, and $R$ is the internuclear separation. Both the initial and final state wave functions depend on the internuclear separation. 

If an external magnetic field is applied, it breaks the degeneracy of molecular hyperfine states and produces a Zeeman shift of the energy levels of the interacting atoms. 
A magnetic Feshbach resonance occurs if the energy of a bound ro-vibrational level in the molecular ground state in closed channel is shifted by the field to coincide with the asymptotic energy of the colliding atoms \cite{RevModPhys.82.1225}.

\begin{figure}[t]
\includegraphics[width=8cm]{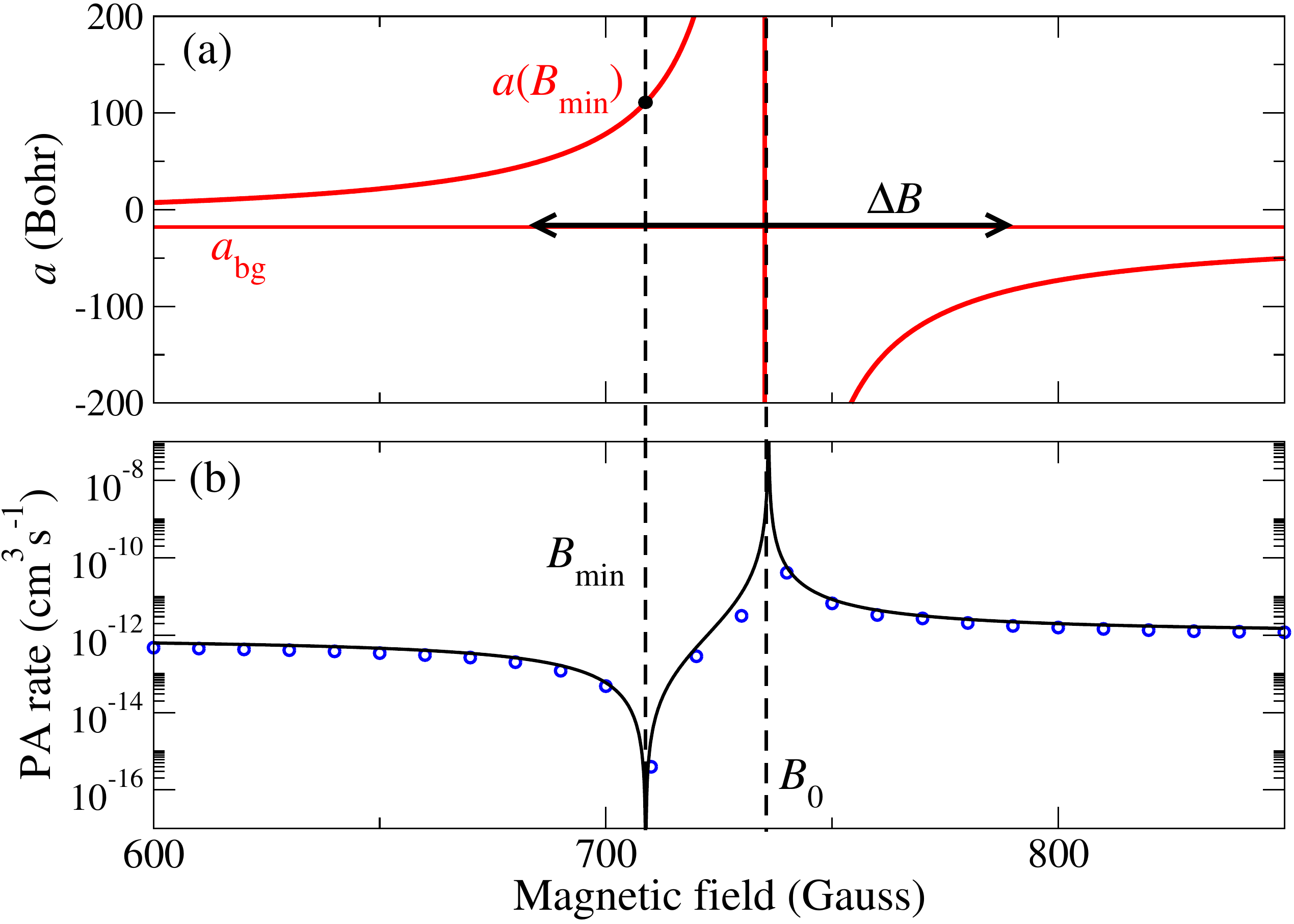}
\caption{Top: Scattering length as a function of magnetic field for diatomic collision of $^7$Li in the initial hyperfine channel $|f_1=1,m_{f_1}=1 \rangle |f_2=1,m_{f_2}=1 \rangle$. The background scattering length $a_{\rm bg}$, the resonance position $B_0$, and the resonance width $\Delta B$ are indicated. Bottom: The PA rate $K_b$ for the formation of $^7$Li$_2$ molecule in $b=(83,1)$ of $1^3\Sigma_g^+$ electronic state for a low laser intensity, $I=1$ mW/cm$^{2}$. Numerical calculation (circles) and two-channel model with $C_1=-7.541$ (solid line) are shown. The position of the sharp minimum at $B_{\rm min}$ and the associated scattering length $a(B_m)\simeq 100$ $a_0$ are also shown.
}
\label{fig1_a_Kpa}
\end{figure}

In the presence of a single magnetic Feshbach resonance, the real part of the atom-atom scattering length $a$ can be parametrized as a function of the magnetic field $B$ according to \cite{2006RvMP...78..483J}
\begin{equation}
   a(B)=a_{\rm bg}(B) \left(1 + \frac{\Delta B}{B - B_0}\right) ,
\label{eq:a(E)}
\end{equation}
where $\Delta B$ is related to the width of the Feshbach resonance (proportional to the coupling strength between the scattering channels), and $B_0$ is the resonant magnetic field. The background scattering length $a_{\rm bg}(B)$ depends on the magnetic field weakly and can be assumed to be constant. A typical behavior of the scattering length $a(B)$ is illustrated in Fig.~\ref{fig1_a_Kpa}(a) for a $s$-wave Feshbach resonance in $^7$Li (see next section for details).

\begin{figure}[t]
\includegraphics[width=\columnwidth]{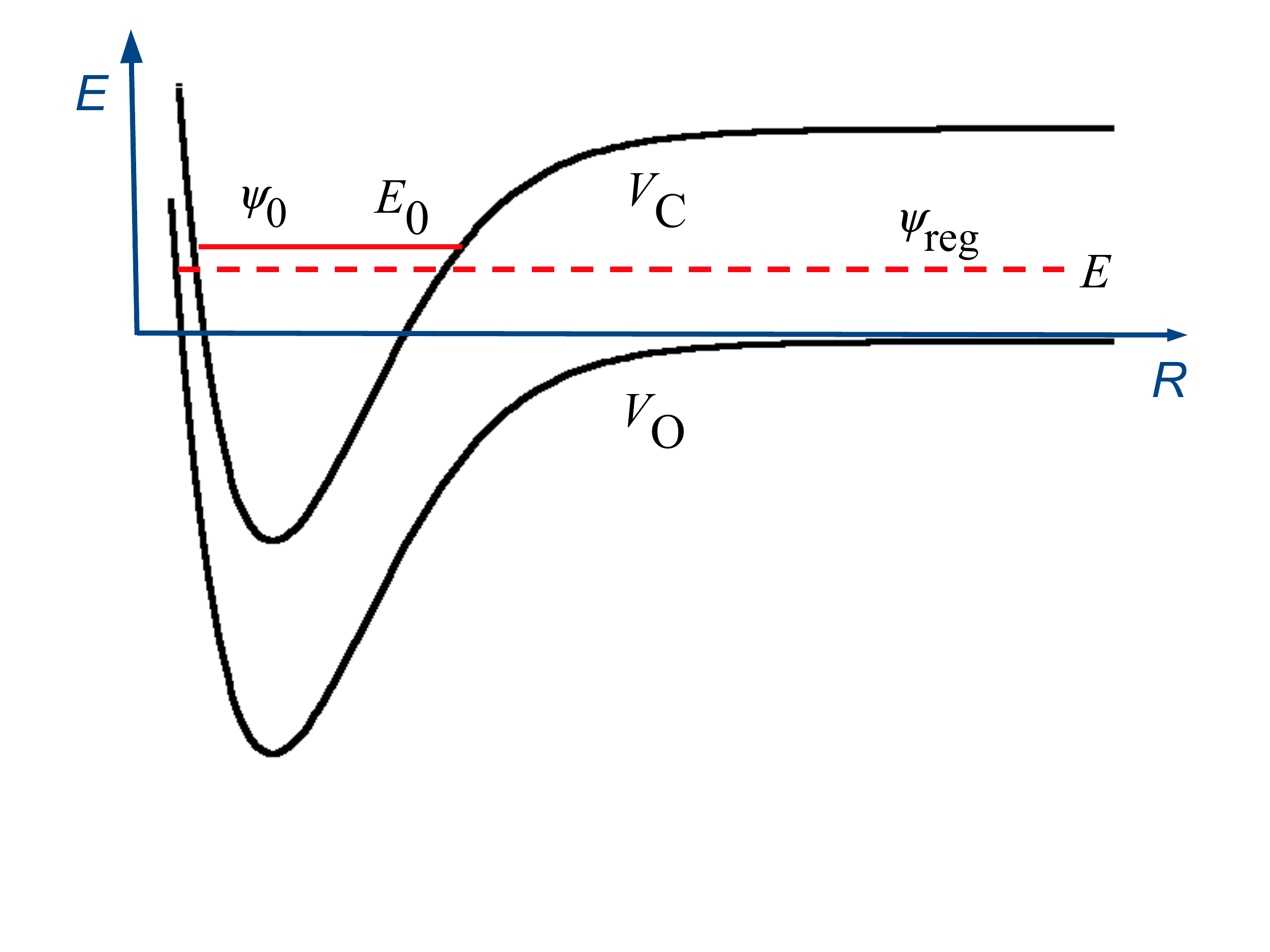}
\caption{Schematic representation of a Feshbach resonance in a two-channel system. The eigenstates, or channels, $|O\rangle$ and $|C\rangle$ correspond to the energeticaly open and closed molecular potentials, $V_O$ and $V_C$, respectively. A Feshbach resonance occurs for the collision energy $E$ equal to the bound state energy $E_0$ in the closed channel.}
\label{fig0}
\end{figure}

Photoassociation in an external magnetic field can be described by a two-channel model that includes one open channel $|\rm O\rangle$ with amplitude $\Psi_{\rm O}(R)$ and a closed channel $|\rm C\rangle$ with amplitude $\Psi_{\rm C}(R)$, typically coupled by hyperfine and Zeeman interactions (Fig. \ref{fig0}) \cite{2008PhRvL.101e3201P}. The total $s$-wave ($\ell =0$) scattering state can then be written as
\begin{equation}
   |\Phi\rangle_{\rm tot} = \Psi_{\rm O}(R) |{\rm O}\rangle + \Psi_{\rm C}(R) |{\rm C}\rangle ,
   \label{eq:Phi-tot}
\end{equation}
with
\begin{eqnarray}
  \Psi_{\rm O} (R) & =& \psi_{\rm reg}(R)
  + \tan \delta \;\psi_{\rm irr}(R) \, \\
  \Psi_{\rm C} (R) & = & -\sqrt{\frac{2}{\pi\Gamma}} \sin \delta \; \psi_0(R) ,
\label{eq:Psi}
\end{eqnarray}
where $\psi_{\rm reg}$ is the ``regular" scattering wave function, {\it i.e.} without coupling to the closed channel, and $\psi_{\rm irr}$ is the ``irregular" scattering wave function arising from that coupling. The amount of coupling is dictated by the resonant phase shift $\delta (k)$, which depends on the proximity of the scattering energy $\varepsilon$ to $E_0$, the energy of a bound state in the closed channel described by the wave function $\psi_0(R)$. Here $\Gamma$ is the width of the resonance. At ultralow energies, the coupling is related to the scattering length by $\tan (\delta + \delta_{\rm bg}) =-ka$, where the background scattering length and phase shift are related by $\tan \delta_{\rm bg} =-ka_{\rm bg}$. 

From Eqs. (\ref{eq:gammas1}) and (\ref{eq:Phi-tot}), the stimulated linewidth $\gamma_s$ can be expressed as \cite{2009NJPh...11e5047P}
\begin{equation}
   \gamma_s = \gamma_{s}^{\rm off} | 1 + C_1 \tan \delta + C_2 \sin \delta |^2 ,
   \label{eq:gammas}
\end{equation}
where $\gamma_s^{\rm off}(\varepsilon,\ell)=(\pi I/\epsilon_0 c) |\langle b | D|\psi_{\rm reg}\rangle|^2$ is the off-resonant stimulated linewidth. The dimensionless coefficients $C_1(b,k)$ and $C_2(b,k)$ are defined for each ro-vibrational level $b$ as ratios of transition dipole matrix elements between the open and closed channel \cite{2008PhRvL.101e3201P}:
\begin{eqnarray}
  C_1(b,k) &=& \frac{\langle b |D| \psi_{\rm irr}\rangle}{\langle b |D| \psi_{\rm reg}\rangle} , \label{eq:C1} \\
  C_2(b,k) &=& -\sqrt{\frac{2}{\pi \Gamma}} \frac{\langle b |D| \psi_0\rangle}{\langle b |D|  \psi_{\rm reg}\rangle}
  \label{eq:C2}  ,
\end{eqnarray}
where $\Gamma$, the true resonance width, is related to $\Delta B$ \cite{friedrich2006theoretical}. 
The relative importance of $C_1$ and $C_2$ depends on the selected target ro-vibrational level $b$. For the molecules produced in high vibrational levels close to the dissociation energy of a particular electronic state, overlap between $|b\rangle$ and $|\psi_0\rangle$ is typically small, leading to $C_2$ much smaller than $C_1$, or $|C_2/ C_1| \ll 1$, so that the second term in Eq. (\ref{eq:gammas}) can be neglected \cite{2008PhRvL.101e3201P}. Therefore, for simplicity, we assume this condition to be valid in the following sections.
Note that $\tan\delta$ and $\gamma_s$ become very large at the resonance, so the saturation effects, even at low laser intensities, become important \cite{2009NJPh...11e5047P}. 

The structure of $\gamma_s$ in Eq. (\ref{eq:gammas}) implies the existence of another feature, shifted away from the resonance (see Fig. \ref{fig1_a_Kpa}); for $\delta \ll 1$, so that $(C_1 + C_2)\delta \simeq -1$, $\gamma_s$ will become small, allowing an approximate expression for the PA rate coefficient $K_b(T)$. In fact, if $\gamma_s/\gamma_b \ll 1$, the $S$-matrix (\ref{Sv}) can then be approximated by \cite{1998PhRvA..58..498C,2006PhRvA..73d1403J}
\begin{equation}
  |S_b(\varepsilon,\ell,\omega)|^2 \approx 2 \pi \gamma_s(\varepsilon,\ell) \delta(\varepsilon - \Delta_b) ,
\end{equation}
and substituting this expression in Eq. (\ref{rate1}), the maximum PA rate is obtained at $\Delta_b=k_B T/2$,
so that a simple expression is obtained
\begin{eqnarray}
  K_b (T) &  = & \frac{2\pi}{h}\frac{e^{-1/2}}{Q_T} \gamma_s(k_B T/2,\ell), \\
  & = & K_b^{\rm off}(T) | 1 + C_1 \tan \delta + C_2 \sin \delta |^2 ,
  \label{K1}
\end{eqnarray}
where $K_b^{\rm off}(T)= \frac{2\pi}{h}\frac{e^{-1/2}}{Q_T} \gamma_s^{\rm off}(k_B T/2,\ell)$. 

Fig.~\ref{fig1_a_Kpa}(b) shows the two features in the PA rate: the large increase near the resonance at $B_0$ and the sharp decrease at $B_{\rm min}$. The validity of the approximate expression (\ref{K1}) away from the resonance is discussed in detail in the next section.

\section{Results and discussion}

\subsection{Photoassociation rate of $^7$Li$_2$}

Photoassociation of $^7$Li$_2$ molecules from an ultracold atomic gas of lithium has been reported to be enhanced by up three orders of magnitude near a Feshbach resonance \cite{PhysRevLett.101.060406,2009NJPh...11e5047P}. Before we proceed to relate the PA rate to the variation of $\beta$, we reproduce the reported results and represent the rate using the two-channel model given in the previous section.

We consider single-photon photoassociation of $^7$Li$_2$ molecules in the bound level $b=(83,1)$ of the $1^3\Sigma_g^+$ electronic state, in the external magnetic field $B$. We also assume that the ultracold atomic gas is initially prepared in the least energetic high-field seeking hyperfine state, represented in the coupled hyperfine basis as $|f_1=1,m_{f_1}=1 \rangle |f_2=1,m_{f_2}=1 \rangle$. In this particular channel there exists a broad Feshbach resonance \cite{RevModPhys.82.1225,Khaykovich17052002} that can be characterized using Eq. (\ref{eq:a(E)}) with $B_0=736$ G, $\Delta B=200$ G and $a_{\rm bg}=-18$ ${\rm a_0}$ (Fig.~\ref{fig1_a_Kpa}(a)).

Using Eq. (\ref{rate1}) with transition dipole moments $D(R)$ from Ref. \cite{1999PhRvA..60.2063C}, we calculate the PA rate $K_b$ for the states $b=(v,1)$, for all vibrational levels $v$ of the electronic state $1^3\Sigma_g^+$ of Li$_2$ molecule. 
We model the dynamics using the Hamiltonian given in Ref. \cite{2008PhRvL.101e3201P} with the spin-spin interaction term neglected, and solve the resulting close-coupling equations. Numerical solution of the multichannel system is obtained using a variable-step mapped Fourier grid method \cite{1999JChPh.110.9865K}, where the box and step size are specified by setting the maximum internuclear distance to $R_{\mathrm{max}}=300$ $a_0$ and mapping potential is calculated based on the local momentum. The coefficients $C_1$ and $C_2$ are determined from the best fit to the form given in Eq. (\ref{eq:gammas}) and compared to the numerical result for a range of PA laser intensities to verify the validity of the two-channel model.
\begin{figure}[t]
\includegraphics[width=8cm]{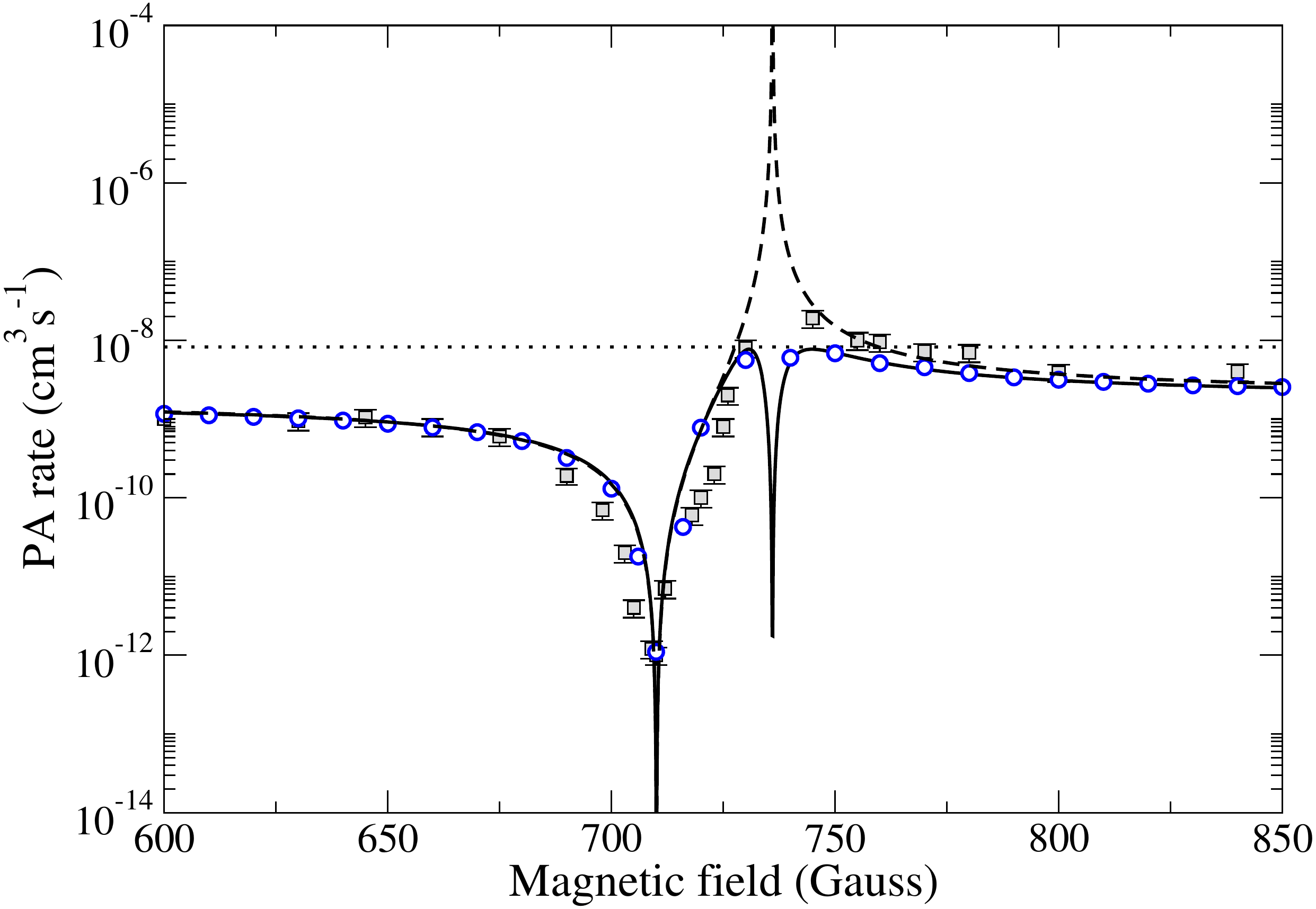}
\caption{Photoassociation rate $K_b$ for $^7$Li$_2$ ($1^3\Sigma_g^+$, $b=(83,1)$) as a function of the magnetic field. The initial hyperfine state $|f_1=1,m_{f_1}=1 \rangle |f_2=1,m_{f_2}=1 \rangle$ is assumed. The results of a full coupled-channel calculation (circles), the two-channel model with $C_1=-7.541$ and $C_2=0$ (solid line), and the experimental measurements \cite{PhysRevLett.101.060406} (squares) with corresponding error bars are shown. The unitarity limit $K_b^{\rm max}$ (dotted line) and the approximate expression for $K_b$ (Eq. (\ref{K1})) (dashed line) are given.}
\label{7Li2_KPA}
\end{figure}
The resulting PA rate for the collision energy $\varepsilon /k_B = 10$ $\mu$K and PA laser intensities $I=1$ mW/cm$^2$ and 1.6 W/cm$^2$, respectively, is shown in Figs.~\ref{fig1_a_Kpa}(b) and \ref{7Li2_KPA}. The conditions in Fig.~\ref{7Li2_KPA} correspond to those reported in the experiment of Hulet's group \cite{PhysRevLett.101.060406}. Our results are in agreement with an earlier theoretical analysis \cite{2009NJPh...11e5047P}. 

The PA rate exhibits two prominent features, at about 710 G and 736 G (Fig. \ref{7Li2_KPA}). The minimum located at $B_{\rm min}$ is caused by the ``destructive interference'' ({\it i.e.} $(C_1+C_2)\delta \simeq -1$), due to small Franck-Condon overlap between the initial scattering state (or Feshbach state) and the target ro-vibrational level $b$.
The second feature appears at the Feshbach resonance ($B_0$), where the initial and final states are resonantly coupled, resulting in a completely saturated transition (the unitarity limit is easily reached). 
At the resonance, according to the two-channel model, $\gamma_b/\gamma_s \ll 1$, resulting in a sharp decrease in $K_b$ \cite{2009NJPh...11e5047P}. 
We note that many phenomena neglected in the two-channel model would need to be included to describe accurately the rate at the resonance, such as a time-dependent treatment to account for the rapid Rabi-flopping between continuum and bound states, and many-body effects due to the strongly interacting nature of the cold gas at resonance (including large three-body recombination rates, {\it etc.}). This is discussed in Refs. \cite{PhysRevLett.101.060406,2009NJPh...11e5047P}.

\subsection{Sensitivity of the photoassociation rate to $\beta$}

\begin{figure}[tb]
   \includegraphics[width=8cm]{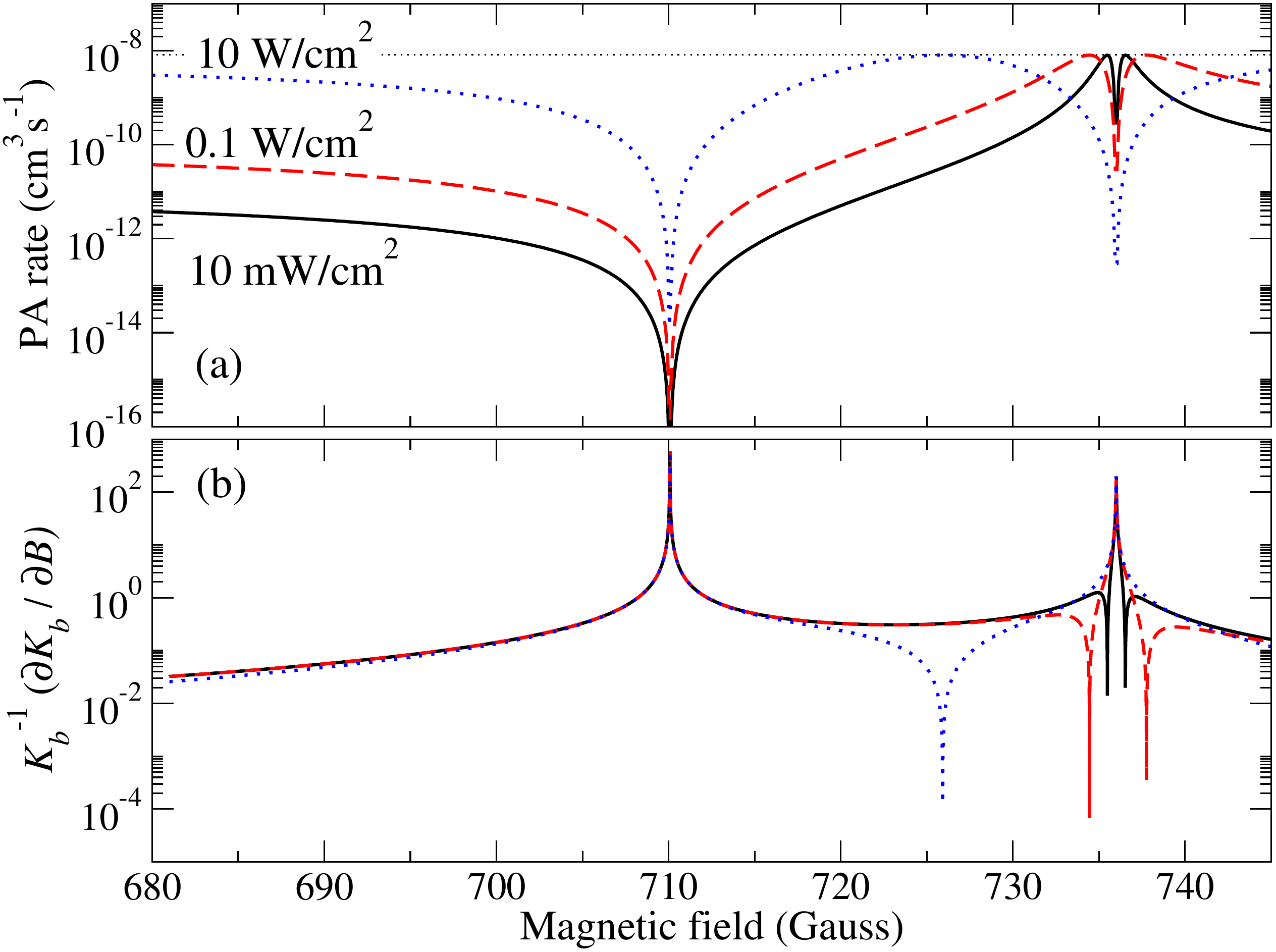}
   \caption{\label{7Li2_intens} Top: Photoassociation rate $K_b$ for $^7$Li$_2$ ($1^3\Sigma_g^+$, $b=(83,1)$) as a function of the magnetic field for three different laser intensities. The unitarity limit is indicated by the dotted line.
   Bottom: Absolute value of the relative variation of the PA rate $K_{b}^{-1} \partial K_b/\partial B$ as a function of the magnetic field.}
\end{figure}

To relate the variation of the PA rate $K_b$ to the variation in $\beta$, we restrict the analysis to the weak-field limit, where $\gamma_s/\gamma_b \ll 1$ and $|S_b(\varepsilon,\ell,\omega)| \ll 1$. As discussed in previous section, the weak-field approximation is generally satisfied for the entire range of magnetic fields, except in the immediate vicinity of the Feshbach resonance. This point is illustrated in the Fig.~\ref{7Li2_intens}(a), where $K_b$ is plotted for three different laser intensities (using the two-channel model to represent $\gamma_s$). Near the resonance at $B_0$, where the PA rate reaches the unitarity limit and saturation effects become relevant, the behavior of $K_b$ depends strongly on the laser intensity. However, near the minimum at $B_{\rm min}$, the rate is not affected by the intensity, aside from a linear scaling factor proportional to the laser intensity $I$ and contained in $\gamma_s^{\rm off}$. This is clearly visible in Fig.~\ref{7Li2_intens}(b), where the dependence on the prefactor is removed by computing $\partial \ln K_b /\partial B$.

Therefore, for simplicity, we focus our treatment on the minimum of the PA rate, $K_b(B_{\rm min})$, where Eq. (\ref{K1}) holds. Furthermore, to obtain analytical expressions, we consider PA to target bound levels $b$ such that $C_2\ll C_1$, which makes it possible for us to neglect the $C_2\sin\delta$ contribution in Eq. (\ref{K1}).

In the ultracold regime, where higher partial waves can be neglected, the $s$-wave phase shift is related to the scattering length as $\tan(\delta + \delta_{\rm bg}) = -k a$, with the background phase shift given by $\tan \delta_{\rm bg} = -k a_{\rm bg}$.
By substituting the phase shift in Eq. (\ref{K1}) with the scattering length, as outlined above, we obtain 
\begin{equation}
    K_b = K_b^{\rm off} \left(1 + C_1 \frac{k (a_{{\rm bg}}-a)}{1 + k^2 a_{{\rm bg}} a}\right)^2 ,
   \label{eq:K}
\end{equation}
where we assume all quantities to be real.
To relate the change in PA rate to variation of $\beta$, we vary Eq. (\ref{eq:K}) with respect to $a$, $a_{\mathrm{bg}}$, and $k$
\begin{equation}
   \delta K_b = \left( \frac{\partial K_b}{\partial a} \right) \delta a + \left( \frac{\partial K_b}{\partial a_{\mathrm{bg}}} \right) \delta a_{\mathrm{bg}} + \left( \frac{\partial K_b}{\partial k} \right) \delta k .
   \label{eq:Kvpart}
\end{equation}
We do not consider the variation of $K_b$ with respect to the coefficient $C_1$ since both dipole matrix elements defining $C_1$ (see Eq. (\ref{eq:C1}) have the same mass dependence leading to a mass-independent coefficient (to first order). We obtain
\begin{eqnarray*}
 \left( \frac{\partial K_b}{\partial a} \right) \delta a & = & -\frac{1+k^2 a_{\rm bg}^2}{1+k^2 a a_{\rm bg}} 
 D(k,a,a_{\rm bg}) a \frac{\delta a}{a} , \\
 \left( \frac{\partial K_b}{\partial a_{\mathrm{bg}}} \right) \delta a_{\mathrm{bg}} \!\! \! & = & \!\! \!    \frac{1+k^2 a^2}{1+k^2 a a_{\rm bg}}  
 D(k,a,a_{\rm bg}) a_{\rm bg} \frac{\delta a_{\rm bg}}{a_{\rm bg}} , \\
  \left( \frac{\partial K_b}{\partial k} \right) \delta k & = &  \frac{1-k^2 a a_{\rm bg}}{1+k^2 a a_{\rm bg}} 
 D(k,a,a_{\rm bg}) (a_{\rm bg}-a)  \frac{\delta k}{k} ,   
 \end{eqnarray*}
where, to simplify the notation, we define
\begin{equation}
 D(k,a,a_{\rm bg}) \equiv \frac{2 K_b C_1 k }{1+C_1 k(a_{\rm bg}-a)+k^2 a a_{\rm bg}} .
\end{equation}
Since we are mainly interested in the region close to the sharp minimum of the PA rate, $K_b(B_{\rm min})$, and $k a \ll 1$ as $k\rightarrow 0$, except at the resonance at $B_0$, we can neglect terms of order $k^2$ in the above equations, leading to
\begin{eqnarray}
 \left( \frac{\partial K_b}{\partial a} \right) \delta a \!\!\! &\simeq & \!\!\! 
 \frac{-2 K_b C_1 k a }{1+C_1 k(a_{\rm bg}-a)} \cdot \frac{\delta a}{a} , \nonumber \\
 \left( \frac{\partial K_b}{\partial a_{\mathrm{bg}}} \right) \delta a_{\mathrm{bg}} \!\!\!& \simeq &\!\!\!   \frac{2 K_b C_1 k a_{\rm bg}}{1+C_1 k(a_{\rm bg}-a)} \cdot \frac{\delta a_{\rm bg}}{a_{\rm bg}}  , \nonumber \\
  \left( \frac{\partial K_b}{\partial k} \right) \delta k \!\!\!&\simeq &\!\!\!  
 \cdot \frac{2 K_b C_1k (a_{\rm bg}-a)}{1+C_1 k(a_{\rm bg}-a)}  \cdot \frac{\delta k}{k} .   
\end{eqnarray}

Chin and Flambaum \cite{PhysRevLett.96.230801} have analyzed the relative variation of the scattering length near a Feshbach resonance using a two-channel model and obtained
\begin{eqnarray}
 \frac{\delta a}{a} & \!\! = \!\!  &\frac{M}{2}\frac{(a-a_{{\rm bg}})^2}{a_{{\rm bg}} a}
        \frac{1}{\rho(E_m)\Delta E}\frac{\delta \beta}{\beta} \equiv \zeta_a \frac{\delta \beta}{\beta} , 
        \label{eq:zeta_a} \\
 \frac{\delta a_{\rm bg}}{a_{\rm bg}}  & \!\! = \!\! & \frac{N\pi}{2} \frac{(a_{\rm bg}-\bar{a})^2+ \bar{a}^2 }
       {a_{\rm bg}\bar{a}} \frac{\delta \beta}{\beta} \equiv \zeta_{a_{\rm bg}} \frac{\delta \beta}{\beta}
       \label{eq:zeta_bg} .
\end{eqnarray}

In Eq. (\ref{eq:zeta_a}), $M$ is the vibrational quantum number of the resonant bound state of energy $E_m$ in the closed channel leading to the resonance of width $\Delta E\propto\Delta B$, and $\rho(E_m)$ is the density of states at the energy $E_m$. In Eq. (\ref{eq:zeta_bg}), $N$ is the the number of bound levels in the open channel (without coupling), and $\bar{a}=c(2\mu C_6/\hbar^2)^{1/4}$ is a mean scattering length (with $c\simeq 0.477 99$) appearing in the background scattering length $a_{\rm bg}=\bar{a}[1-\tan (\phi -\pi/8)]$, where the phase $\phi$ is obtained from a semi-classical treatment \cite{PhysRevLett.96.230801}. Finally, noting that $k=\sqrt{2\mu \varepsilon/\hbar^2}=\mbox{const.}\times \beta^{1/2}$, we find $\delta k/k = \delta\beta/2 \beta$. 
Grouping all those terms, we obtain (near the minimum)
\begin{eqnarray}
 \left. \frac{\delta K_b}{K_b} \right|_{\rm min} & = & \frac{-2 C_1 k a }{1+C_1 k(a_{\rm bg}-a)}
 \bigg\{ \zeta_a - \frac{a_{\rm bg}}{a} \zeta_{a_{\rm bg}} \nonumber \\
 &&  +\frac{(a- a_{\rm bg})}{2a}\bigg\} \frac{\delta \beta}{\beta}
 \label{eq:K_all}
\end{eqnarray}
As discussed above, at the minimum of the PA rate the scattering length $a(B_{\rm min})$ is finite and well behaved, {\it i.e.} unlike at $B_0$ where $a\rightarrow \pm \infty$, or at $B=B_0-\Delta B$ where $a=0$, according to Eq. (\ref{eq:a(E)}). Therefore, $|a|\sim a_{\rm bg}$ near $B_{\rm min}$, implying that the last term in the bracket of Eq. (\ref{eq:K_all}) is of the order of unity, while the second term is of the order $\zeta_{\rm bg}$. Also, unless an accidental ``zero-energy resonance'' is present for a particular system, $\bar{a} \simeq a_{\rm bg}$, and the number of bound level $N$ is at most of the order $10^2$, and so is $\zeta_{\rm bg}\sim N\pi/2$. The coefficient $\zeta_a$, however, can become very large. 
As pointed out in Ref. \cite{PhysRevLett.96.230801}, $\zeta_a$ exhibits three enhancement factors that can lead to a large enhancement: the vibrational quantum number $M$, the factor affected by the scattering length near the resonance, and the third term that contains the resonance width $\Delta E$, since $\rho(E_m)\Delta E$ can be very small. Therefore, for the sake of simplicity, we keep only the first term and write
\begin{equation}
   \label{eq:K2}
 \left. \frac{\delta K_b}{K_b} \right|_{\rm min} = \frac{-2 C_1 k a }{1+C_1 k(a_{\rm bg}-a)} \zeta_a  
 \frac{\delta \beta}{\beta} \equiv \xi_b \frac{\delta \beta}{\beta}
\end{equation}
We refer to $\xi_b$ as the \textit{enhancement factor}, and direct comparison with Eq. (\ref{eq:zeta_a}) shows that we have gained an extra term on $\delta a/a$ obtained by Chin and Flambaum \cite{PhysRevLett.96.230801}, which depends on $C_1$, $a$ and $a_{\rm bg}$. Using the expression (\ref{eq:a(E)}) for $a$, we rewrite the enhancement factor as
\begin{eqnarray}
  \xi_b & = & \frac{M C_1 k }{[C_1 k(a-a_{\rm bg})-1]}\frac{(a-a_{{\rm bg}})^2}{a_{\rm bg}\rho(E_m)\Delta E} \\
  &=& M \frac{a_{\rm bg} C_1 k}{(B-B_0)(B_{\rm min}-B)} 
       \frac{\Delta B}{\alpha\rho(E_m)}
       \label{eq:L2}
\end{eqnarray}
where $B_{\rm min} = B_0-a_{\rm bg} C_1 k \Delta B$, and we used $\Delta E = \alpha\Delta B$ with $\alpha = h$ 840 MHz/G for Li$_2$ \cite{2009NJPh...11e5048C}. The result in Eq. (\ref{eq:L2}) shows that for a given resonance, $\xi_b$ will diverge for two values of the magnetic field $B$; first at the resonance $B_0$ (where the above treatment is not accurate), and at $B_{\rm min}$ which coincides with the \textit{minimum} of the PA rate observed in \cite{2008PhRvL.101e3201P, PhysRevLett.101.060406}.

So, by measuring the variation of $K_b$, an enhanced measurement of $\delta\beta$ could be achieved. As hinted by Eq. (\ref{eq:L2}) and expected from the large increase in $K_b$ near the resonance \cite{2008PhRvL.101e3201P, PhysRevLett.101.060406}, one can anticipate such an enhancement at the resonance. However, even a low-intensity PA laser will fully saturate the transition at the Feshbach resonance and prevent a consistent measurement ({\it e.g.}, due to many-body interactions and large three-body decay). In contrast, the minimum of the PA rate for $B=B_{\rm min}$ is not affected by the saturation effects and its relative variation does not change with laser intensity, as shown in Fig. \ref{7Li2_intens}. Therefore, it should be possible to increase the PA laser intensity at the minimum to obtain higher molecular formation rate in experimental realizations.

The enhancement factor $\xi_b$ varies with both the properties of the Feshbach resonance and the target bound state $b$ considered. In particular, the position of $B_{\rm min}$ changes for different target level $b$, as has been observed in earlier studies \cite{2009NJPh...11e5047P,0953-4075-42-19-195202}. By selecting the optimal target state $b$, one can improve significantly the detection threshold based on the relative variation of the scattering length.
Finally, the temperature of the ultracold gas determines the collisional energy, possibly affecting the profile and positions of Feshbach resonances, even though these effects can be neglected in a sufficiently cold gas.

\begin{figure}[tb]
\includegraphics[width=8cm]{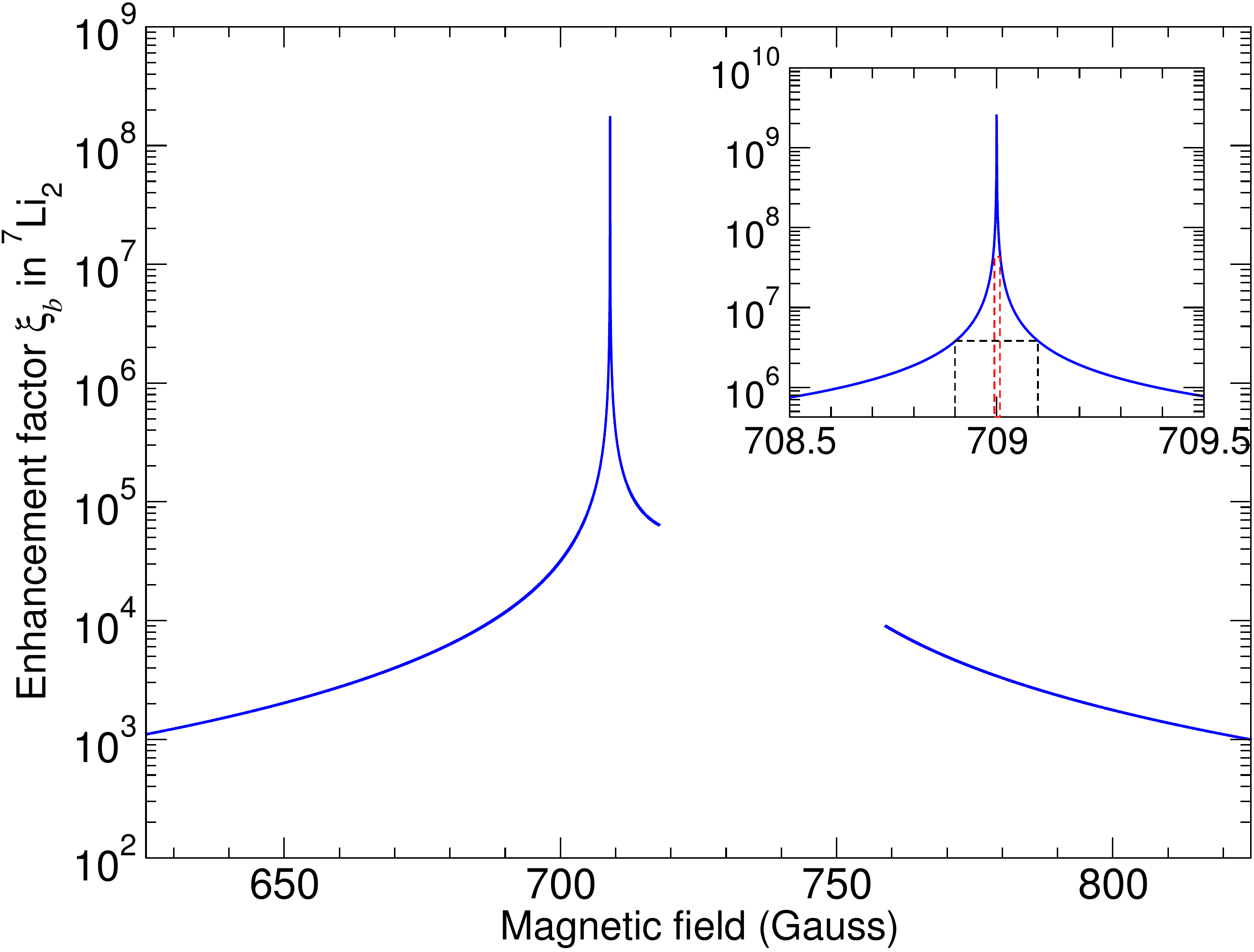}
\caption{\label{7Li2}The enhancement factor $\xi_b$ into $b=(83,1)$ of $^7$Li$_2$ (same initial and final states as above). The line is discontinued in the region where the expression (\ref{eq:L2}) is not valid due to the saturation effects. \textit{Inset:} Zoom in on the peak. The enhancement factor for the magnetic field measured up to 0.1 G and 0.01 G is indicated by black and red dashed lines, respectively.}
\end{figure}

\subsection{Detection limits from Li$_2$ and LiNa molecules}

We proceed to obtain an estimate of the attainable sensitivity of the photoassociation rate on the variation of $\beta$. The enhancement factor $\xi_b$ in $^7$Li$_2$ was calculated for the ro-vibrational level $b=(83,1)$ of $1^3\Sigma_g^+$ molecular state formed as described above. We focus on the region around the minimum of the PA rate at $B=710$ G where Eq. (\ref{eq:L2}) holds. The resulting enhancement as a function of the magnetic field is shown in Fig. \ref{7Li2}. 

To better quantify the total number of molecules formed, we assume a density of an ultracold gas in the trap to be $n_{\rm Li} \approx 10^{12}$ ${\rm cm}^{-3}$, a PA laser intensity of $I=1$ W/cm$^2$, and a volume illuminated by the PA laser of $V = 1$ mm$^3$. At $B=710.09$ G, the PA rate is $K_b = 9.4 \times 10^{-18}$ cm$^3$s$^{-1}$, resulting in the enhancement factor $\xi_b \approx 2 \times 10^8$. If the total number of molecules formed can be detected with an accuracy of 1$\%$, detecting about 100 out of $10^4$ molecules while keeping the magnetic field stable to about 0.01 G at 710 G, the variation of $\beta$ up to $5 \times 10^{-11}$ could be detected. 
In genereal, the total number of photoassociated molecules can be increased by using higher PA laser intensity, while a better control of the magnetic field lowers the detection threshold.

\begin{figure}[t]
   \includegraphics[width=8cm]{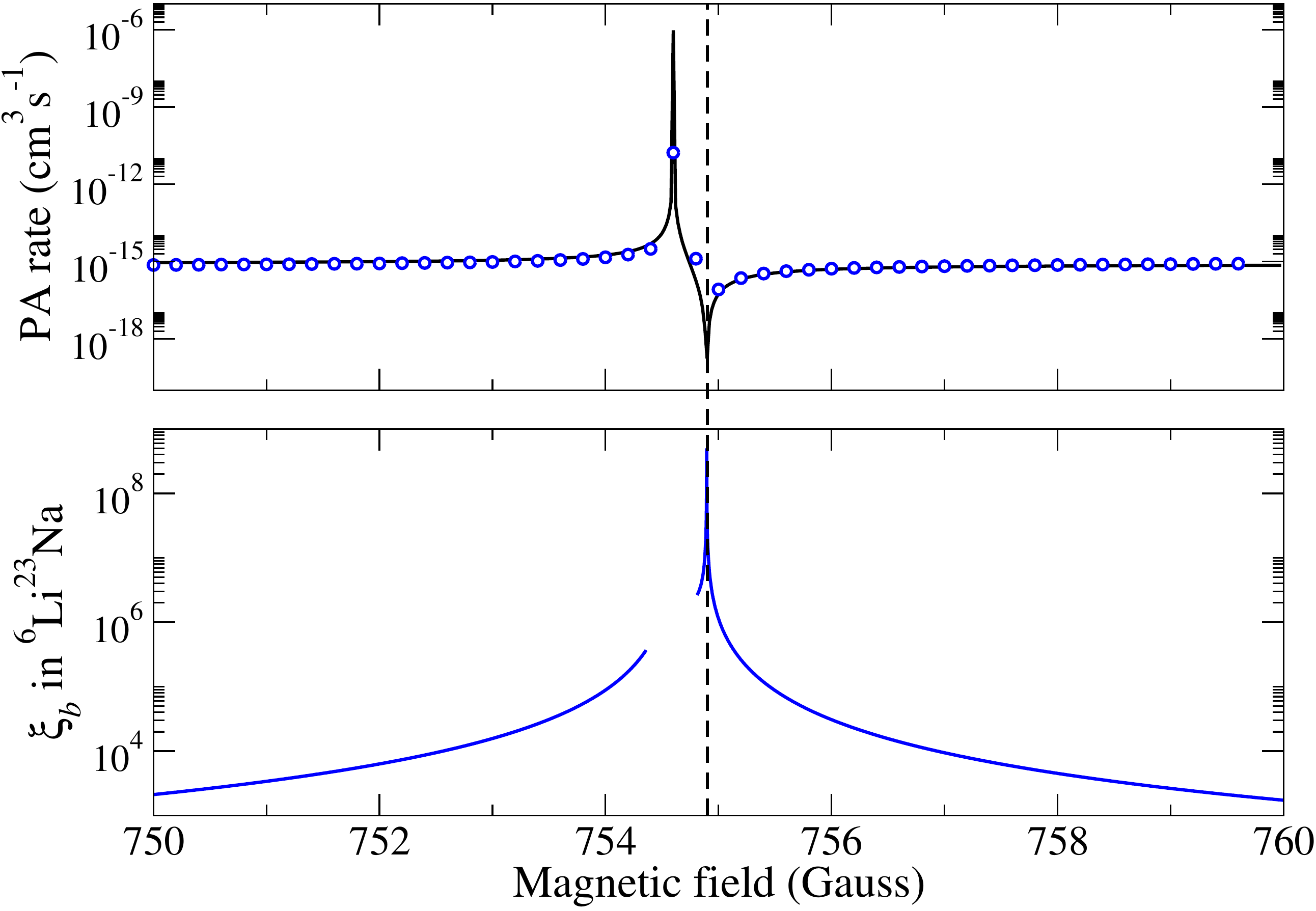}
   \caption{\label{LiNa_fig} \textit{Top:} Photoassociation rate $K_b$ for $^6$Li$^{23}$Na ($1^1\Pi$, $v=59,j=1$) for the PA laser intensity $I=1$ W/cm$^{2}$ as a function of the magnetic field.
   \textit{Bottom:} The enhancement factor $\xi_b$ for the corresponding magnetic field.}
\end{figure}

As pointed out in Ref. \cite{PhysRevLett.96.230801}, a better test of temporal variation of $\beta$ could be obtained using a narrower Feshbach resonance in a heavy molecule. To illustrate the impact of the resonance width on the enhancement factor and detection sensitivity, we consider a magnetic Feshbach resonance found at about $B = 759.0$ G in the least energetic initial hyperfine channel $|11 \rangle$ in $^6$Li$^{23}$Na. This resonance was originally assigned as an $s$-wave resonance \cite{PhysRevLett.93.143001,2008PhRvA..78a0701G}, and recently re-assigned as a very narrow $d$-wave resonance \cite{PhysRevA.85.042721}.
Using the same approach as outlined above for Li$_2$, we calculate the enhancement factor $\xi_b$ for forming $^6$Li$^{23}$Na molecules in the ro-vibrational state $b$ of the $1^1\Pi$ electronic state. As before, we solve the resulting coupled-channel equations in a magnetic field, using the most recent potential energy curves to model the two lowest-energy singlet and triplet states \cite{PhysRevA.85.042720}, as well as the \textit{ab-initio} potential energy curve for the excited $1^1\Pi$ state, and the corresponding transition dipole moments \cite{2008JPhB...41o5101M}, while the spin-orbit interaction and other non-Born-Oppenheimer corrections were not included in the model. The resulting Franck-Condon factors are particularly favorable for forming molecules in the ro-vibrational level $b=(59,1)$. We show the corresponding PA rate $K_b$ and the enhancement factor $\xi_b$ in Fig.~\ref{LiNa_fig}. 

We find that the PA rate of $^6$Li$^{23}$Na molecules within the two-channel model can be parametrized by $a_{\mathrm{bg}}=-73$ $a_0$, $\Delta B = 0.02$ G, $B_0=759.0$ G \cite{PhysRevA.85.042721}, and $C_1=-480$, $K_{b}^{{\rm off}}=4 \times 10^{-11}$ cm$^3$s$^{-1}$. The last vibrational level in the $(X)1 ^1\Sigma^+$ state gives $M=44$ \cite{PhysRevA.85.042720}. 
The minimum of the PA rate occurs at about $B=754.9$ G, where we predict that about $1.8\times 10^5$ molecules per second can be formed for the PA laser intensity of 1 W/cm$^2$. 
Assuming gas density and illuminated volume as in the calculation for Li$_2$, and the stability of the magnetic field of up to $10^{-3}$ G at the resonance, this corresponds to the enhancement factor $\xi_{b=(59,1)} \approx 10^9$. Hence, observing the variation of the PA rate on the order of 10 molecules per second would amount to detecting a change in $\beta$ of about 1 in $10^{-13}$. 

\section{Summary}

In the present study, we developed a simple connection between variation of the photoassociation rate of diatomic molecules near a magnetic Feshbach resonance and variation of the electron-proton mass ratio $\delta \beta / \beta$ based on the work of Chin and Flambaum \cite{PhysRevLett.96.230801}. Our theory is based on the demonstrated sharp variation of the photoassociation rate for a diatomic pair near a Feshbach resonance. While the expresions derived in the paper are specific for photoassociation of diatomic molecules near a magnetic Feshbach resonance, they could be generalized to other systems with similar properties, such as molecular ions or Rydberg molecules.
The derived proportionality relation was used to estimate the sensitivity of the photoassociation rate to possible temporal or spatial variations of $\beta$.
The enhanced sensitivity of the photoassociation rate to the variation of $\beta$ was demonstrated in ultracold Li and Li-Na gases, where we made an attempt to develop a realistic model and obtain reasonable estimates of the maximum attainable detection threshold under experimental conditions.

Our calculations indicate that it would be possible to detect relative variability of $\beta$ on the order of 1 in $10^{-13}$ by monitoring the PA rate of ultracold LiNa molecule.
At the moment, several existing and proposed precision measurement experiments aimed at detecting variation of $\beta$ are able to achieve higher sensitivity. However, our estimates are rather conservative with respect to the experimental parameters and based on currently available and well-studied gases. We expect a significant increase in detection sensitivity could in heavier diatomic molecules, such as Yb$_2$ or $^{133}$Cs$_2$ molecules, photoassociated near a very narrow Feshbach resonance.
For example, a much narrower optical Feshbach resonance with an externally controlled coupling strength  \cite{2004PhRvL..93l3001T} or a $g$-wave magnetic Feshbach resonance in $^{133}$Cs \cite{2003Sci...301.1510H}, would increase the enhancement factor by several orders of magnitude. 
Assuming the resonance width of $\Delta B = 5$ mG \cite{2003Sci...301.1510H}, and the detection efficiency of molecules of 0.1 $\%$ near the minimum of the PA rate, a test of variation of $\beta$ on the $10^{-15}-10^{-16}$ level could be performed. This is comparable to the most accurate current laboratory measurements of $\beta$. Finally, a number of narrow Feshbach resonances has recently been observed at low magnetic fields in heavy systems such as erbium and dysprosium \cite{2012PhRvL.108u0401A,2013arXiv1312.6401B}. Such systems could allow an even better control over the magnetic field and possibly further improve the detection threshold of the proposed approach.


We thank P. Pellegrini and I. Simbotin for stimulating discussions. The work of M.G. was partially supported by the Army Research Office Grant No. W911NF-13-1-0213 (63925CH Chemistry Division), and the work of R.C. by the National Science Foundation Grant No. PHY 1101254.


\bibliography{varKPA}



%
\end{document}